\documentclass[iop,onecolumn]{emulateapj}
\usepackage[utf8]{inputenc}
\usepackage[english]{babel}
\usepackage{natbib}
\usepackage{amsmath}
\usepackage{comment}
\usepackage{graphicx}
\begin {document}

\newcommand{\hatpone} {HAT-P-1}
\newcommand{\hatponeb}{HAT-P-1b}
\newcommand{\hdegy}   {HD 189733}
\newcommand{\hdegyb}  {HD 189733b}

\newcommand{\hdkettob}{HD 209458b}
\newcommand{\ads}     {ADS 16402}
\newcommand{\Teq}     {\ensuremath{T_\mathrm{eq}}}

\title {Relative photometry of \hatponeb{} occultations}
\author {Bence Béky\altaffilmark{1},
Matthew J.~Holman\altaffilmark{1},
Ronald L.~Gilliland\altaffilmark{2},
Gáspár Á.~Bakos\altaffilmark{3,4,5},
Joshua N.~Winn\altaffilmark{6},
Robert W.~Noyes\altaffilmark{1},
Dimitar D.~Sasselov\altaffilmark{1}}
\shortauthors {B\'{e}ky et al.}
\altaffiltext {1} {Harvard--Smithsonian Center for Astrophysics, 60 Garden St, Cambridge, MA 02138, USA}
\altaffiltext {2} {Center for Exoplanets and Habitable Worlds, The Pennsylvania State University, 525 Davey Laboratory, University Park, PA 16802, USA}
\altaffiltext {3} {Department of Astrophysical Sciences, Princeton University, 4 Ivy Lane, Princeton, NJ 08544, USA}
\altaffiltext {4} {Alfred P.~Sloan Research Fellow}
\altaffiltext {5} {Packard Fellow}
\altaffiltext {6} {Department of Physics, and Kavli Institute for Astrophysics and Space Research, Massachusetts Institute of Technology, 70 Vassar Street, Cambridge, MA 02139, USA}
\email {bbeky@cfa.harvard.edu}

\begin {abstract}

We present HST STIS observations of two occultations of the transiting exoplanet \hatponeb{}. By measuring the planet to star flux ratio near opposition, we constrain the geometric albedo of the planet, which is strongly linked to its atmospheric temperature gradient. 
An advantage of \hatpone{} as a target is its binary companion \ads{} A, which provides an excellent photometric reference, simplifying the usual steps in removing instrumental artifacts from HST time-series photometry. We find that without this reference star, we would need to detrend the lightcurve with the time of the exposures as well as the first three powers of HST orbital phase, and this would introduce a strong bias in the results for the albedo. However, with this reference star, we only need to detrend the data with the time of the exposures to achieve the same per-point scatter, therefore we can avoid most of the bias associated with detrending. Our final result is a $2\sigma$ upper limit of 0.64 for the geometric albedo of \hatponeb{} between 577 and 947 nm. 

\end {abstract}


\section {Introduction}

The effective temperature of Jovian extrasolar planets in close orbits is strongly influenced by irradiation from their host stars. It is the Bond albedo, defined as the reflected fraction of incident electromagnetic power, that determines how much of this irradiation contributes to the thermal balance of the planet. Atmospheric temperature, in turn, determines -- for a given composition -- the presence and position of absorbers, clouds and other structures determining the albedo. Once the albedo is inferred from observations, the goal is to find a self-consistent atmospheric model and temperature.

Even though it is the Bond albedo that can be used directly in the calculation of effective temperature, the geometric albedo $A_\mathrm g$ is more accessible observationally. It is defined as the ratio of reflected flux at opposition (zero phase angle) to the reflected flux by a hypothetical flat, fully reflecting, diffusely scattering (Lambertian) surface of the same cross-section at the same position. The geometric albedo of Lambertian surfaces is at most one: for example, a fully reflecting Lambertian sphere has a geometric albedo of $\frac23$ \citep {1992essi.book.....H}. However, some surfaces reflect light preferentially in the direction where it came from (a phenomenon known as opposition surge), and thus can exhibit geometric albedos exceeding one \citep[e.g.][]{1990Icar...88..407H}.

A simple way to measure albedo is to observe an occultation (secondary eclipse) of a transiting exoplanet and directly compare the brightness of the star only (while the planet is occulted) to the total brightness of the star and the planet near opposition (shortly before or after the occultation). Examples of other phenomena that can be used to constrain albedo but are not discussed in this paper are phase variations \citep[e.g.][]{2006Sci...314..623H}, Doppler-shift of reflected starlight \citep[e.g.][]{2007arXiv0711.2304L}, and polarization of reflected starlight \citep[e.g.][]{2006PASP..118.1302H}.

To mention specific examples, the exoplanet \hdkettob{} has mass $M=0.69\;M_\mathrm J$, radius $R=1.36\;R_\mathrm J$, orbital period $P=3.52$ days, and zero-albedo equilibrium temperature $\Teq=1450$ K \citep{2008ApJ...677.1324T}. When calculating \Teq, perfect heat redistribution on the planetary surface is assumed. \citet {2008ApJ...689.1345R,2009IAUS..253..121R} observe the occultation of \hdkettob{} in the 400--700 nm bandpass with the Microvariability and Oscillations of Stars (MOST) satellite. They find a geometric albedo of $A_\mathrm g=0.038\pm0.045$, and infer a $1\sigma$ upper limit of 0.12 on the Bond albedo, indicating the absence of reflective clouds. Based on atmospheric models, this constrains the atmospheric temperature to between 1400 K and 1650 K. Normally, a cloud-free atmosphere exhibits low albedo due to the strong pressure-broadened absorption lines of neutral sodium and potassium \citep {2000ApJ...538..885S}. However, Spitzer Space Telescope observations of \hdkettob{} at 3.6, 4.5, 5.8, and 8.0 $\mu$m indicate water emission features, suggesting temperature inversion in the higher atmosphere, which hints that an unknown absorber is present at low pressure \citep {2008ApJ...673..526K,2007ApJ...668L.171B}.

Another well-studied example is \hdegy{} \citep[$M=1.14\;M_\mathrm J$, $R=1.14\;R_\mathrm J$, $P=2.22$ days, $T_\mathrm{eq}=1200$ K,][]{2008ApJ...677.1324T}. \citet {2008Natur.456..767G} and \citet{2008ApJ...686.1341C} observe its occultations with the Spitzer Space Telescope. They find strong water absorption features, indicating the lack of temperature inversion in the atmosphere. \citet {2011MNRAS.416.1443S} perform transmission spectroscopy on this planet, and interpret the results as indicating high altitude haze. This would cause the planet to exhibit high geometric albedo in the visible. \citet {2009IAUS..253..121R} report on MOST observations of this planet's occultation, but cannot constrain the albedo due to the high activity of the host star. Note that \hdkettob{} and \hdegyb{} only differ by a few hundred kelvins in terms of \Teq, yet seem to exhibit very different atmospheres.

The subject of this work is \hatponeb{}, a transiting exoplanet with mass $M=0.524\;M_\mathrm J$, radius $R=1.225\;R_\mathrm J$, orbital period $P=4.47$ days \citep{2008ApJ...686..649J}, and zero-albedo equilibrium temperature $\Teq=1300$ K \citep {2008ApJ...677.1324T}. This last value is between those of \hdkettob{} and \hdegyb{}. Indeed, \citet {2010ApJ...708..498T} observe two occultations of \hatponeb{} with Spitzer and infer a modest temperature inversion in the atmosphere from the occultation depths at 3.6, 4.5, 5.8, and 8.0 $\mu$m. In the light of these observations, constraining the geometric albedo of this planet in the visible and near infrared would be useful for better understanding its atmospheric structure, and for refining atmospheric models.

In this paper, we report on observations of two occultations of \hatponeb{}. Section \ref {sec:observations} presents the details of the observations, flux extraction, and detrending. We describe how we calculate the geometric albedo in Section \ref {sec:albedo}, with special attention to handling each uncertainty source, and comparing relative photometry results to those without using the reference star. We discuss our findings in Section \ref {sec:discussion}.


\section {Observations and data analysis}
\label {sec:observations}

The exoplanetary host star \hatpone{} is a member of the wide binary system \ads, with $11.2"$ projected separation at a distance of 139 pc. \hatpone, also known as \ads{} B, with $V=10.4$ is only 0.4 magnitude fainter than its binary companion \ads{} A of the same G0V spectral type. This allows for relative (differential) photometry to mitigate the effect of systematic errors.

\subsection {Observations and data preprocessing}
A proposal was accepted as GO 11069 to observe two occultations (secondary eclipses) of \hatponeb{} with the Hubble Space Telescope (HST) Advanced Camera for Surveys (ACS) High Resolution Channel (HRC). However, ACS failed in 2007 January, before the observations would have been carried out, and HRC remains inoperational to date. Instead, another HST instrument, the Space Telescope Imaging Spectrograph (STIS) carried out the observations, as program GO 11617. Two occultations of \hatponeb{} were observed during two visits, including two orbits before and one after each occultation. Because of the brightness of the targets, spectroscopy was required to allow for reasonably long exposures. To capture both stars without very tight constraints on spacecraft orientation, we did not use a slit. Table \ref {tab:observations} summerizes the details of the observations. With these settings, the largest electron count in each exposure was about half the well size, well below saturation. However, it is interesting to note that longer exposures would not have posed a problem in terms of linearity either: when STIS pixels get saturated, the excees charge bleeds into surrounding pixels with virtually no loss, and summing these pixel counts still results in a linear response \citep {1999PASP..111.1009G}.

\begin {deluxetable}{lcc}
\tabletypesize {\scriptsize}
\tablecaption {HST/STIS program GO 11617 observation parameters\label {tab:observations}}
\tablehead {& visit 1 & visit 2}
\startdata
HJD at beginning of first exposure & 2\,455\,544.8269 & 2\,455\,888.6602 \\
HJD at end of last exposure & 2\,455\,545.1181 & 2\,455\,888.9497 \\
number of orbits & 5 & 5 \\
number of spectra for each orbit & 19+23+23+23+23 & 19+23+23+23+23 \\
number of spectra in total & 111 & 111 \\
grating & \multicolumn2c{\texttt{G750L}} \\
slit & \multicolumn2c{slitless} \\
exposure time & \multicolumn2c{100 s} \\
cadence & \multicolumn2c{128 s} \\
subarray size & \multicolumn2c{380x1024 pixels} \\
gain & \multicolumn2c{4}
\enddata
\end {deluxetable}

\begin {figure}
\begin {center}
\includegraphics*[width=89mm]{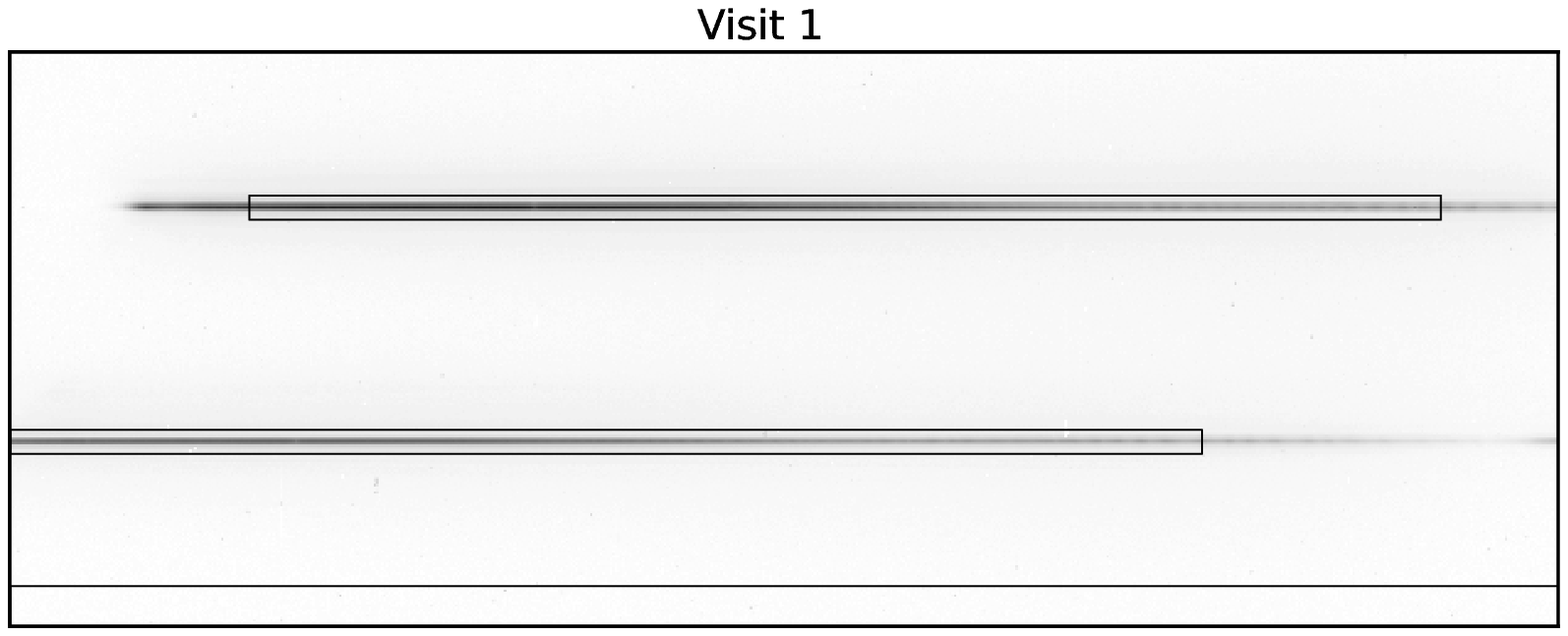}
\includegraphics*[width=89mm]{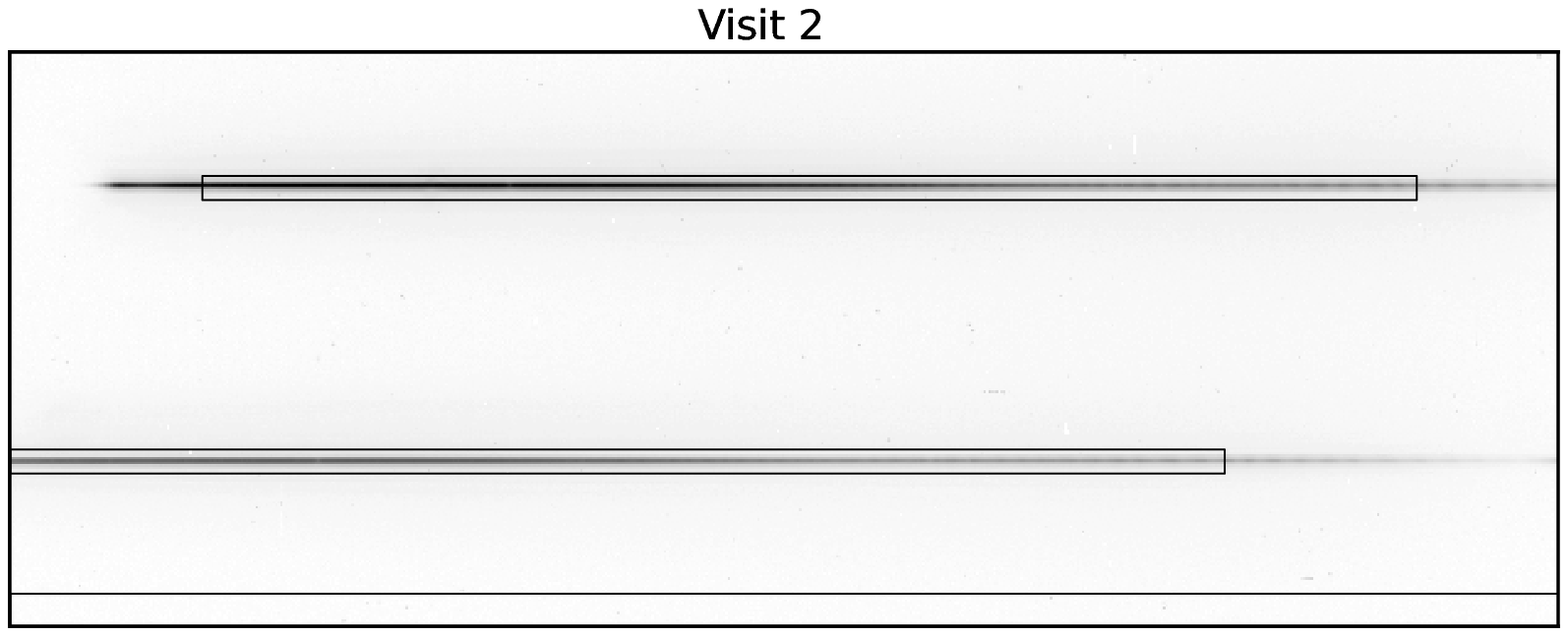}
\end {center}
\caption {The first spectrum used in the final analysis (the second exposure of the second orbit) of the first and second visits (left and right panels, respectively). The upper star is \ads{} A, the lower one is the planetary host \ads{} B. The rectangular apertures around the spectra are also shown. The bottom rectangle is the background aperture. Note the different cross-dispersion distance and dispersion direction shift between the two stars' spectra for the two visits due to spacecraft orientation. The blue end of the stellar apertures is determined by where the detector edge cuts the specrum of \ads{} B.}
\label {fig:spectra}
\end {figure}

We identify hot pixels and exclude them from our apertures, and identify cosmic rays and substitute them by the average of the previous and next frame values. Then we perform rectangular aperture photometry on the two stars, and define an aperture in the entire length of the detector in the dispersion direction for sky background estimation. For each exposure, we subtract the background aperture photon count from the stellar aperture photon counts, scaled by the number of pixels. Figure \ref {fig:spectra} shows a typical exposure from each visit, with the apertures used. The blue end of the stellar apertures is 564 nm for the first visit and 557 nm for the second. The difference is caused by different orientations of the telescope around its optical axis, resulting in the STIS detector edge cutting the \hatpone{} spectrum at different positions for the two visits. These wavelength values are calculated after identifying the H$\alpha$ and Na I D lines in the stellar spectra. We extract the same spectral range for the two stars within a visit to fight wavelength-dependent systematics. However, we allow for different blue end cuts between visits, otherwise we would lose too many photons in the second visit due to the more restrictive wavelength limit of the first one. We do not expect the geometric albedo to vary significantly due to this small change in blue end wavelength cut.

\subsection {Detrending}

The next step is to detrend the data, that is, to mitigate instrumental effects by subtracting multiples of vectors describing circumstances of the observations. We try detrending with time (to remove the overall linear trend within a visit), HST orbital phase and its powers (to remove orbitwise periodic variations), CCD housing temperature (CCD chip temperature is not available with the current Side-2 electronics), a focus model provided by HST Observatory Support, fine pointing data available from telemetry, and fine pointing data based on a fit for the position of the spectra on the CCD. To detrend, we simultaneously fit for free parameters of the lightcurve model (reference flux and planet-to-star flux ratio) as well as detrending vector coefficients using a linear algebraic least square method. We actually use the magnitude of \hatpone{} or the magnitude difference of the two stars, that is, we assume that instrumental effects are multiplicative. Since both the planet-to-star flux ratio and detrending corrections are very small, this is equivalent to assuming additive effects and fitting for the flux or flux ratio. Each detrending vector is mean subtracted so that they do not change the average stellar magnitude. To quantify the effect of detrending and avoid overfitting, we minimize the Bayesian Information Criterion (BIC), which is the sum of $\chi^2$ and a term penalizing extra model parameters \citep {Schwarz1978}.

When analyzing STIS data to perform photometry on exoplanetary host stars, \citet {2007ApJ...655..564K} find it justified to detrend with a cubic polynomial of HST orbital phase, whereas \citet {2001ApJ...552..699B} and \citet {2011MNRAS.416.1443S} use fourth order polynomials. Indeed, if we only consider the lightcurve of \hatpone, we find the lowest BIC when detrending with time at mid-exposure and the first three powers of HST orbital phase. However, if we divide the lightcurve of the planetary host by that of the reference star ADS 16402 A, we find the lowest BIC when detrending only with mid-exposure time. This shows that relative photometry is less sensitive to systematics, and can mitigate HST orbital effects enough that detrending with orbital phase is not justified. The Bayesian Information Criterion does not justify detrending with the temperature, focus, or jitter vectors in either case.

Figure \ref {fig:lightcurves} presents the raw lightcurves (panels a--d), without detrending, which indeed show strong orbitwise periodic variations. Panels (e, f) show the lightcurve of \hatpone{} detrended with time at mid-exposure and the first three powers of HST phase, demonstrating that this procedure corrects for the overall linear trend and most of the orbitwise periodic variation. On the other hand, the raw relative lightcurves shown on panels (g, h) do not exhibit such large variations, and we only need to remove the linear trend (panels i, j). These panels all have logaritmic vertical axes with the same scaling, so that relative scatter is directly comparable. Note that since we perform a simultaneous fit of the occultation lightcurve and detrending vectors, the resulting $\chi^2$ tells us how close the data are to a model accounting for both the occultation and systematics, without the danger of misinterpreting the occultation as scatter.

The strength of these observations is the presence of the reference star, which already proves to be advantageous. In order to quantify how much it improves the albedo limits, we perform a full analysis both without and with this reference data, independently tuning all extraction parameters.

\subsection {Aperture parameters and data omission}

If we only extract the flux of \hatpone, we find that we obtain the least scatter after detrending (with mid-exposure time and first three powers of HST orbital phase) if the stellar apertures are 14 and 16 pixels wide in cross-dispersion direction for the two visits, respectively, and the stellar spectra are cut off at 788 nm. Redward of this wavelength there is extra scatter due to fringing, which is difficult to combat in slitless mode. If we use the flux ratio of the two stars, we get the least scatter after detrending (this time with mid-exposure time only) if the stellar apertures are 16 and 18 pixels wide in cross-dispersion direction for the two visits respectively, and the stellar spectra are cut off further in the near infrared at 947 nm. Using the reference star thus allows us to extract photons from a larger aperture. The optimal background aperture is 27 pixels wide for the first visit and 22 for the second in both cases.

We estimate sky background using an aperture placed as far from the stellar spectra as possible, to avoid contamination by starlight. Varying the background aperture width by a few tens of pixels introduces scatter of 0.02 in the best fit albedo as long as the aperture is not too narrow and not too close to the stars either. Given the other uncertainty sources discussed in Section \ref {sec:uncertainty}, this means that the results are practically insensitive to the exact choice of the background aperture. However, if we place the background aperture between the two spectra on the detector, or expand it on the side to get within 50 pixels of \hatpone, we find a larger average error, of 0.05, in the geometric albedo due to stray starlight.

\begin {figure}
\begin {center}
\includegraphics*[width=89mm]{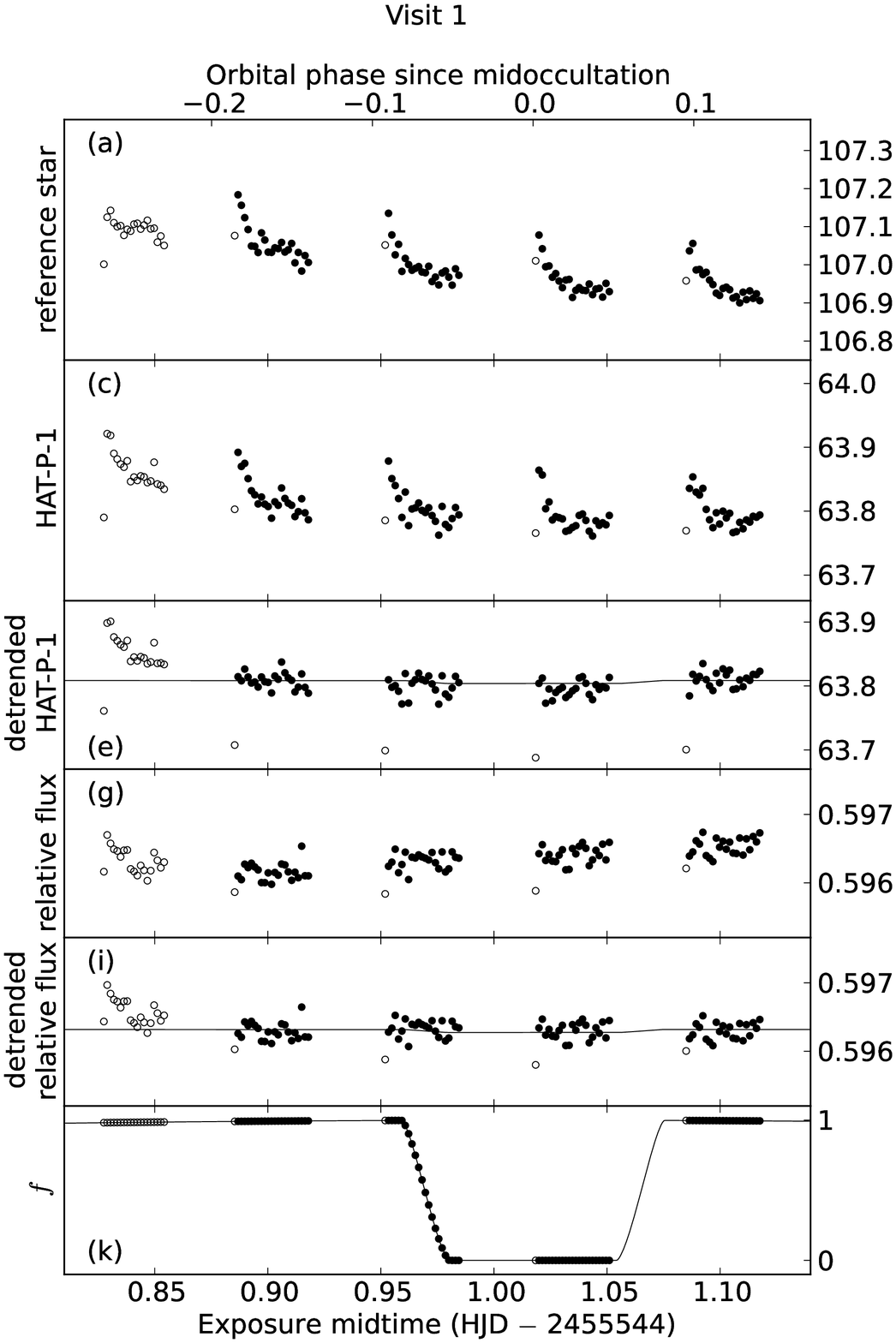}
\includegraphics*[width=89mm]{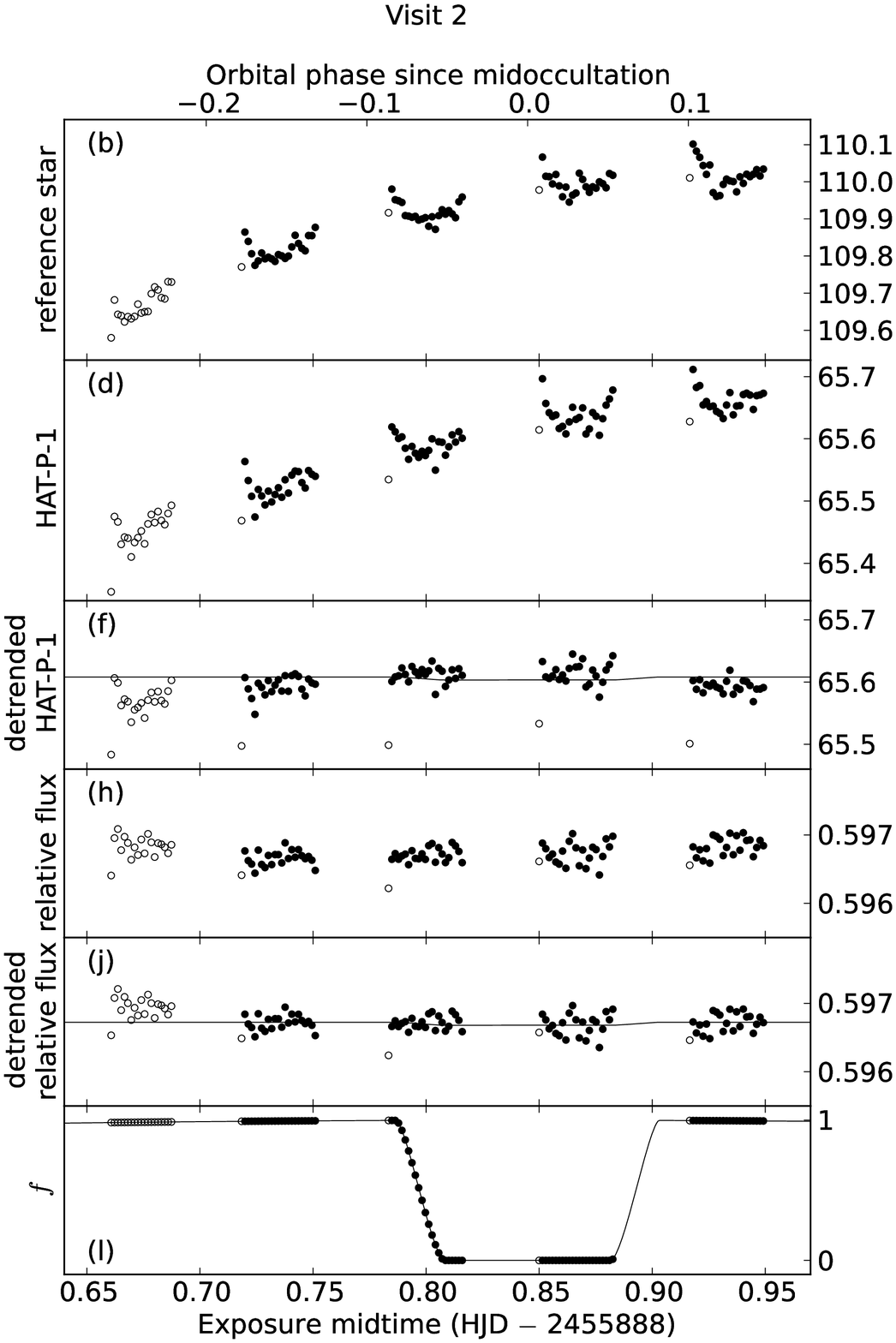}
\end {center}
\caption {Panels (a, b): background-subtracted photon count per exposure for reference star \ads{} A, in million photons. Panels (c, d): same for planetary host star \hatpone. Panels (e, f): photon count of \hatpone{} detrended with time at mid-exposure and first three powers of HST orbital phase, in million photons. The model lightcurve of a fully reflecting planet is overplotted with a solid line. Panels (g, h): relative flux, that is, background-subtracted photon count of \hatpone{} divided by background-subtracted photon count of reference star. Panels (i, j): relative flux detrended with time at mid-exposure. The model lightcurve of a fully reflecting planet is overplotted with a solid line. Panels (k, l): the fraction $f$ of the planetary surface that is illuminated and unobscured, with the values for each exposure overlaid on the continuous curve. The bottom horizontal axes show mid-exposure time in HJD, the top horizontal axes show $\phi-\phi_0$, the planetary orbital phase since midoccultation. Panels (a--j) have logarithmic vertical axes with the same scaling, so that relative scatter is directly comparable. Panels (a), (c), (e), (g), (i) and (k) display the first visit, (b), (d), (f), (h), (j), and (l) the second. Filled circles represent data points included in the analysis, empty circles the omitted ones.}
\label {fig:lightcurves}
\end {figure}

Finally, we investigate whether it is justified to omit data points. For example, \citet {2011MNRAS.416.1443S} and \citet {2007ApJ...655..564K}
both omit the first orbit of each five orbit visit, also the first exposure of each subsequent orbit, because these data points exhibit larger scatter. The scatter of the first orbit might be attributed to the thermal settling of the spacecraft after its new pointing. We therefore calculate the scatter per data point for all twelve possible combinations of omitting the first orbit or not, omitting the first or first two exposures of each orbit or not, and omitting the last exposure of each orbit or not. Comparing these results, we find that both for the analysis of \hatponeb{} only and for relative photometry using the reference star, it justified to omit the first orbit of each visit and the first exposure of each subsequent orbit, but not more, consistently with \citet {2011MNRAS.416.1443S} and \citet {2007ApJ...655..564K}. We therefore keep 88 data points per visit. These data points are represented with filled circles on Figure \ref {fig:lightcurves}, whereas omitted data points are represented with empty ones.

Note that the data analysis parameter space has many dimensions: stellar and background aperture geometry, data omission, and choice of detrending parameters. The choices presented above were found after multiple iterations, and they yield the minimum residual sum of squares in each dimension, but strictly speaking, they might not represent the global minimum.

\section {Upper limit on geometric albedo}
\label {sec:albedo}

\subsection {Lightcurve model}

To model the occultation lightcurve, we adopt the transit ephemeris, the planetary radius and orbital semi-major axis in stellar radius units, and transit impact parameter values and uncertainties reported by \citet {2008ApJ...686..649J}. We assume that the orbit is circular, consistently with theoretical expectations, radial velocity measurements \citep {2007ApJ...656..552B}, and occultation timing \citep {2010ApJ...708..498T}. We account for the light travel time of 55 s across the planetary orbit, and neglect the thermal radiation of the planet. We calculate the projected area of the unobscured part of the dayside of the planet by assuming that both the stellar and planetary disks are circular, and the terminator line on the planet is an arc of an ellipse. Let $f$ denote this area relative to the total planetary disk. Then the observed flux at any given time is $F_\star + fF_\mathrm p$, where $F_\star$ is the stellar flux, and $F_\mathrm p$ is the flux of the planet in opposition. Note, however, that we can never observe the theoretical maximum flux $F_\mathrm p$ from the planet, because opposition happens during occultation (therefore $f$ is always smaller than one). We derive the following model for $f$:
\begin {align}
\nonumber
d &= \sqrt {R_\star^2 b^2 + a^2 \sin^2 (\phi-\phi_0)}, \quad\quad
u = \arccos\frac {d^2 + R_\star^2 - R_\mathrm p^2}{2dR_\star}, \quad\quad
v = \arccos\frac {d^2 + R_\mathrm p^2 - R_\star^2}{2d R_\mathrm p} \\
f &=  \begin {cases}
0 & \textrm {if } d \leqslant R_\star - R_\mathrm p \textrm { (occultation),} \\
\max \left( \frac {1+\sqrt{1-\frac{d^2}{a^2}}} 2 - \left(\frac u\pi - \frac{\sin2u}{2\pi}\right)\frac{R_\star^2}{R_\mathrm p^2} - \left(\frac v\pi - \frac{\sin2v}{2\pi}\right), 0 \right) & \textrm {if } R_\star - R_\mathrm p < d \leqslant R_\star + R_\mathrm p \textrm { (ingress and egress),} \\
\frac {1+\sqrt{1-\frac{d^2}{a^2}}} 2 & \textrm {if } R_\star + R_\mathrm p < d \textrm { (out of occultation).}
\end {cases}
\end {align}
Here 
$R_\star$ is the stellar radius,
$R_\mathrm p$ is the planetary radius,
$a$ is the planetary orbital radius, 
$\phi$ is the planetary orbital phase,
$\phi_0$ is the orbital phase at midoccultation, 
$b$ is the impact parameter, 
$d$ is the projected distance of the center of the planetary and stellar disks, 
and $u, v \in [0,\pi]$ are auxiliary functions for the case of ingress and egress.
The bottom panels of Figure \ref {fig:lightcurves} show $f$ as a continuous function of time, with the value in the middle of each exposure overlaid.

This model gives the exact result as long as the boundary of the stellar disk and the terminator line on the planet do not intersect in projection, and is extended in a monotonic and continuous manner at the end of ingress and beginning of egress when they do. This introduces a small error which only affects a few data points. Note, however, that there are other sources of error: when determining the terminator line, we assume that the planet is irradiated by parallel rays from the direction of the center of the star, instead of properly calculating irradiation in the belt from where the star is partially seen on the horizon of the planet. Also, we do not account for that the substellar point is closer to the star, therefore it is exposed to more irradiation. These errors are in the order of $\frac{R_\star^2}{a^2}\approx0.01$, that is, they bias our geometric albedo estimate by the order of 1\%.

We assume Lambertian reflectance of the incident light. Note that the maximum angle of stellar irradiance on the planet to the line of sight during our science exposures used in the final analysis is $\approx11^\circ$. In case of reflection from materials like regolith, this would be very different from reflection in opposition, but we expect the reflected flux for a gas giant like \hatponeb{} to depend only weakly on the incident angle.

Using this model lightcurve, we then fit simultaneously for the planet to star flux ratio at opposition and the coefficients of the detrending vectors. Then the geometric albedo can be calculated using the expression \citep[e.g.][]{2008ApJ...689.1345R}
\begin {equation}
A_\mathrm g = \frac {F_\mathrm p}{F_\star} \frac{a^2}{R_\mathrm p^2}. 
\end {equation}
To illustrate the magnitude of the effect with respect to the scatter of the data, Figure \ref {fig:lightcurves} panels (e, f) and (i, j) feature a plot of a model lightcurve, assuming that the planet is a fully reflecting Lambertian sphere with a geometric albedo of $\frac23$.

\subsection {Uncertainty sources}
\label {sec:uncertainty}

There are two sources of errors in the inferred geometric albedo: observational errors and contribution of parametric uncertainties of the system. We estimate the total observational uncertainty $u_\mathrm{obs}$ by bootstrapping. Since instrument parameters like temperature might not have a normal distrubtion or might not influence the data in a linear fashion, we cannot assume that the error distribution is normal. Therefore we apply the bias corrected accelerated bootstrap method \citep {1987JASA..82..171}, a generalized bootstrap algorithm that partially corrects for effects due to non-normal error distributions.

The total observational uncertainty is due to photon noise and other uncertainties due to instrument thermal instability, pointing jitter, and stellar activity in both stars. (Quantization noise due to the gain set to 4 can be neglected.) We estimate the uncertainty $u_\mathrm{photon}$ due to photon noise by independently redrawing every data point (\hatpone, reference star, and background for each exposure) from a Poisson distribution with a parameter given by the original photon count, and recalculating the geometric albedo. Finally, we get the estimate of the uncertainty $u_\mathrm{other}$ due to other sources by subtracting the photon noise estimate from the total observational uncertainty estimate in quadrature.

When quantifying the uncertainty contributions of the planetary system parameters, we divide them into two groups: ephemeris (mid-transit time and period), and geometry (planetary radius and orbital semi-major axis relative to the stellar radius, and impact parameter). This division is important for two reasons: first, the two visits took place 1181 and 1525 days after the reference mid-transit ephemeris of \citet {2008ApJ...686..649J}, respectively, therefore we expect that the second visit will suffer more from the uncertainty in mid-occultation time. Second, it is easier to refine the ephemeris by photometric follow-up transit observations than to refine the geometric parameters, so we want to assess how much this would improve the results. Note that geometric parameters not only influence the occultation lightcurve shape that we use for fitting, but also factor in when converting the planet-to-star flux ratio to geometric albedo. We estimate the uncertainties $u_\mathrm{ephem}$ in the geometric albedo due to uncertainties in ephemeris, $u_\mathrm{geom}$ due to those in geometric parameters, and their total contribution $u_\mathrm{param}$ due to all parametric uncertainties by redrawing the respective parameters from independent normal distributions defined by their best fit values and uncertainties, and recalculating the geometric albedo in each case. We confirm that $u_\mathrm{ephem}$ and $u_\mathrm{geom}$ add up in quadrature to $u_\mathrm{param}$. We use 1\,000\,000 bootstrap iterations to estimate the observational uncertainty $u_\mathrm{obs}$, and 1\,000\,000 random drawings to estimate each of $u_\mathrm{photon}$, $u_\mathrm{ephem}$, $u_\mathrm{geom}$, and $u_\mathrm{param}$.

Finally, we add the observational and parametric uncertainty estimates in quadrature to estimate the total uncertainty $u_\mathrm{total}$, and add this to the best fit geometric albedo to calculate the upper limits. The breakdown of uncertainty sources is represented in a tree structure in Tables \ref{tab:only} and \ref{tab:relative}: the upper limit is the sum of its children nodes best fit and total uncertainty, and uncertainties are quadrature sums of their children nodes.

We find that most lower albedo limits of the final analysis are negative, therefore we only present upper limits for the geometric albedo. The $1\sigma$ and $2\sigma$ upper limits are determined so that the probability of the geometric albedo being smaller than this is $0.6827$ and $0.9545$, respectively. As a comparison, the one-sided $1\sigma$ and $2\sigma$ upper limits of a normal distribution occur at $0.475\sigma$ and $1.690\sigma$ above the mean, respectively, and we expect our corresponding uncertainties to have a similar ratio. 

Table \ref{tab:only} presents the best fit values, uncertainties due to various phenomena, and upper limits for the geometric albedo, based on the lightcurve of \hatpone{} only. We performed the calculations for the two visits separately, and also for a joint model, where a single geometric albedo value was fit for the two visits, but we allowed for different coefficients of the detrending vectors for the two visits. The upper limits highlight the weakness of detrending: the $2\sigma$ upper limit based on the first visit data is meaningless (greater than one), and even worse, the limit based on the second visit data only is unphysical (negative). The reason for this is that the detrending vectors are not orthogonal to the occultation signal, that is, systematic effects have a component that mimics the occultation lightcurve. Therefore even though detrending is justified by the Bayesian Information Criterion, it introduces a bias in the geometric albedo values.

On the other hand, if we feed the flux ratio of the planetary host star \hatpone{} and the reference star \ads{} A into the occultation lightcurve model, much less detrending is justified, thus we expect less bias in the result. Indeed, Table \ref{tab:relative} shows that the results from the two visits are much closer to each other, and all upper limits are positive. Even though the uncertainties in the two cases are very similar, as we can tell by comparing values in Tables \ref{tab:only} and \ref{tab:relative}, in the second case, this is achieved by using only one detrending vector instead of four. This shows the enormous advantage of the reference star: to diminish the need for detrending, therefore arrive at similar uncertainties with much less bias. We adopt the upper limits of the joint fit as our final result.

\begin {deluxetable*} {llllcccccc}
\tabletypesize {\scriptsize}
\tablecaption {Best fit values, uncertainties due to different sources, and upper limits for the geometric albedo, based on \hatpone{} lightcurve only, for comparison\label{tab:only}}
\tablehead {\; & \; & \; & \; & \multicolumn2c {visit 1} & \multicolumn2c {visit 2} & \multicolumn2c {joint fit} \\
&&&& $1\sigma$ & $2\sigma$ & $1\sigma$ & $2\sigma$ & $1\sigma$ & $2\sigma$}
\startdata
\multicolumn4l {$A_g$ upper limit} & 1.44 & 1.94 & $-1.99$ & $-1.46$ & $-0.38$ & 0.03 \\
& \multicolumn3{|@{}l} {-- best fit $A_g$} & \multicolumn2c {$1.25$} & \multicolumn2c {$-2.19$} & \multicolumn2c {$-0.54$} \\
& \multicolumn3{|@{}l} {-- $u_\mathrm{total}$} & 0.19 & 0.69 & $\phantom{-}0.21$ & $\phantom{-}0.74$ & $\phantom{-}0.16$ & 0.57 \\
&& \multicolumn2{|@{}l} {-- $u_\mathrm{obs}$} & 0.19 & 0.68 & $\phantom{-}0.20$ & $\phantom{-}0.72$ & $\phantom{-}0.16$ & 0.57 \\
&& \multicolumn1{|l}{\;} & \multicolumn1{|@{}l} {-- $u_\mathrm{photon}$} & 0.15 & 0.52 & $\phantom{-}0.14$ & $\phantom{-}0.51$ & $\phantom{-}0.10$ & 0.36 \\
&& \multicolumn1{|l}{\;} & \multicolumn1{|@{}l} {-- $u_\mathrm{other}$} & 0.12 & 0.44 & $\phantom{-}0.14$ & $\phantom{-}0.51$ & $\phantom{-}0.12$ & 0.44 \\
&& \multicolumn2{|@{}l} {-- $u_\mathrm{param}$} & 0.03 & 0.11 & $\phantom{-}0.05$ & $\phantom{-}0.17$ & $\phantom{-}0.01$ & 0.05 \\
&&& \multicolumn1{|@{}l} {-- $u_\mathrm{ephem}$} & 0.00 & 0.00 & $\phantom{-}0.03$ & $\phantom{-}0.08$ & $\phantom{-}0.01$ & 0.04 \\
&&& \multicolumn1{|@{}l} {-- $u_\mathrm{geom}$} & 0.03 & 0.11 & $\phantom{-}0.04$ & $\phantom{-}0.15$ & $\phantom{-}0.01$ & 0.03
\enddata
\end {deluxetable*}

\begin {deluxetable*} {llllcccccc}
\tabletypesize {\scriptsize}
\tablecaption {Best fit values, uncertainties due to different sources, and upper limits for the geometric albedo, based on the lightcurves of \hatpone{} and reference star \ads{} A, our final results\label{tab:relative}}
\tablehead {\; & \; & \; & \; & \multicolumn2c {visit 1} & \multicolumn2c {visit 2} & \multicolumn2c {\textbf{joint fit}} \\
&&&& $1\sigma$ & $2\sigma$ & $1\sigma$ & $2\sigma$ & $\mathbf1\pmb\sigma$ & $\mathbf2\pmb\sigma$}
\startdata
\multicolumn4l {$A_g$ upper limit} & 0.08 & 0.59 & 0.55 & 1.14 & $\mathbf{0.24}$ & $\mathbf{0.64}$ \\
& \multicolumn3{|@{}l} {-- best fit $A_g$} & \multicolumn2c {$-0.13$} & \multicolumn2c {$0.31$} & \multicolumn2c {$\mathbf{0.09}$} \\
& \multicolumn3{|@{}l} {-- $u_\mathrm{total}$} & 0.20 & 0.71 & 0.23 & 0.83 & 0.15 & 0.54 \\
&& \multicolumn2{|@{}l} {-- $u_\mathrm{obs}$} & 0.19 & 0.71 & 0.23 & 0.83 & 0.15 & 0.53 \\
&& \multicolumn1{|l}{\;} & \multicolumn1{|@{}l} {-- $u_\mathrm{photon}$} & 0.15 & 0.57 & 0.16 & 0.56 & 0.11 & 0.39 \\
&& \multicolumn1{|l}{\;} & \multicolumn1{|@{}l} {-- $u_\mathrm{other}$} & 0.12 & 0.42 & 0.17 & 0.61 & 0.10 & 0.36 \\
&& \multicolumn2{|@{}l} {-- $u_\mathrm{param}$} & 0.03 & 0.10 & 0.01 & 0.07 & 0.01 & 0.07 \\
&&& \multicolumn1{|@{}l} {-- $u_\mathrm{ephem}$} & 0.02 & 0.08 & 0.00 & 0.05 & 0.01 & 0.05 \\
&&& \multicolumn1{|@{}l} {-- $u_\mathrm{geom}$} & 0.01 & 0.05 & 0.01 & 0.06 & 0.01 & 0.05
\enddata
\end {deluxetable*}

By comparing the different contributions to the uncertainty of the geometric albedo, we see that the observational uncertainties are much larger than the parametric ones. This means that performing further photometric observations of occultations could significantly improve the albedo upper limit, whereas using additional transit observations to refine the ephemeris and geometric parameters and use them to reanalyze these data would not. As for the observational uncertainties, residual systematic uncertainty $u_\mathrm{other}$ and photon noise contribution $u_\mathrm{photon}$ are within a factor of two, which tells us that further improving our data analysis could push the upper albedo limits down only by a small amount. Surprisingly, we do not always find $u_\mathrm{ephem}$ to be larger for the second visit than for the first, but we confirm that $1\sigma$ and $2\sigma$ uncertainties of each kind have a ratio close to what is expected for a normal distribution.

\section {Discussion and summary}
\label {sec:discussion}

Based on HST STIS observations, we established $0.24$ as the $1\sigma$, and $0.64$ as the $2\sigma$ upper limit for the geometric albedo of \hatponeb{} in the 557--947 nm band. Unfortunately, this limit is not tight enough to determine whether there is temperature inversion in the atmosphere. This question is relevant because \hatponeb{} has an equilibrium temperature between that of \hdkettob{} (thought to exhibit temperature inversion) and \hdegyb{} (thought not to). In addition, a better constrained albedo would provide information about the actual atmospheric temperature of the planet, as well as indicate the presence or absence of reflective clouds or high-altitude haze.

Our data analysis demonstrates that the reference star \ads{} A helps us greatly to reduce systematic effects. Even though detrending with powers of HST orbital phase would equally reduce scatter in the signal \citep[just like demonstrated by][]{2001ApJ...552..699B,2007ApJ...655..564K,2011MNRAS.416.1443S}, we find that it introduces a bias in the geometric albedo estimate. We attribute this effect to the fact that these detrending vectors are not orthogonal to the occultation signal. Most of this bias can be avoided by performing relative photometry, so that much less detrending is necessary to mitigate systematic effects. This is possible because \ads{} A has a similar brightness and same spectral type as \hatpone{}, and their angular distance is small enough so that they fit in the STIS field of view, but large enough so that their PSFs do not overlap.

We found that the uncertainties of the system parameters have a negligible effect on the geometric albedo uncertainty. The dominant uncertainty sources are photon noise and other noise effects (thermal instability, pointing jitter, other systematics, and astrophysical noise of the two stars), the contributions of which scale inversely with the square root of the number of observations. This also means that additional observations would improve the upper limit, without being too limited by how precisely we know the geometry and ephemeris of the planetary system. For example, if the geometric albedo of \hatponeb{} was 0.1, then approximately seven times more observations would be required to arrive at 0.4 as a $3\sigma$ upper limit, enough to infer the absence of an omnipresent reflective cloud layer.

It is interesting to note that these observations were originally proposed for the grism instrument ACS/HRC, which has a total throughput of 0.15--0.25 in this wavelength range, as opposed to 0.04--0.08 for STIS with the \texttt{G750L} grating. 
Thus ACS/HRC observations would have resulted in roughly three times more photons. Assuming the same $u_\mathrm{other}$ and $u_\mathrm{param}$ values, this approximately translates to a $u_\mathrm{total}$ of $0.12$ instead of $0.15$ for the $1\sigma$ limit, and $0.43$ instead of $0.54$ for $2\sigma$.

The binary companion star can help data analysis of further \hatpone{} observations (e.g., Wakeford et al., submitted). Also, similar methods could be used for other planetary hosts in binary systems. A suitable example is XO-2, of magnitude $V=11.2$, with the companion XO-2 S of $V=11.1$ at 31" separation. This companion star has been used as a reference for transmission spectroscopy both from the ground with GTC \citep {2012MNRAS.426.1663S}, and with HST NICMOS \citep {2012ApJ...761....7C}. XO-2 is a potential target for relative photometry during occultation with HST STIS in slitless mode: the $P=2.6$ day orbital period of XO-2b \citep {2007ApJ...671.2115B} would mean a larger planet-to-flux ratio than in case of \hatponeb{} for the same geometric albedo, and its zero-albedo equilibrium temperature $\Teq=1300$ K \citep {2008ApJ...677.1324T}, similar to that of \hatponeb{}, would make such a measurement interesting in terms of atmospheric models.

\acknowledgements
M.J.H. and J.N.W. gratefully acknowledge support from NASA Origins grant NNX09AB33G.
G.Á.B. acknowledges support from NSF grant AST-1108686 and NASA grant NNX12AH91H.

\bigskip

\bibliography {b}

\end {document}